\documentclass[aps,prd,preprint,superscriptaddress,showpacs,longbibliography,
footnoteadded]{revtex4-1}

\usepackage[T1]{fontenc}
\usepackage{float}
\usepackage{subfigure}
\usepackage{amssymb}
\usepackage{graphicx}
\usepackage{amsmath}
\usepackage{slashed}
\usepackage{tensor}
\usepackage{hyperref}
\usepackage{epstopdf}
\usepackage{extarrows}
\usepackage{color}
\usepackage{bbm}
\usepackage[section]{placeins}
\usepackage{makecell}

\usepackage[utf8]{inputenc}
\hypersetup{
    colorlinks=true,
    linkcolor=red,
    citecolor=blue,
}

\begin{document}

\title{Quasinormal modes  of phantom Reissner-Nordström-de Sitter black holes}
\author{Hang Liu}
\email{hangliu@mail.nankai.edu.cn}
\affiliation{College of Physics and Materials Science, Tianjin Normal University, Tianjin 300387, China}

\begin{abstract}
In this paper, we investigate some characteristics of phantom Reissner-Nordström-de Sitter (RN-dS) black holes. The peculiar features of phantom field  render this kind of black holes quite different from their counterparts. We can only find at most two horizons in this spacetime, i.e. event horizon and cosmological horizon. For the black hole charge parameter, we find that it is not bounded from below.  We calculate quasinormal modes (QNMs) frequencies  of massless neutral scalar field perturbation  in this black hole spacetime, and some properties related to the large charge parameter are disclosed.  
\end{abstract}


\maketitle

\section{Introduction}
One of the stunning discoveries of the modern physics is the observation of the accelerating expansion of our Universe. To understand the underlying physics of this profound phenomena, people have devoted large amounts of endeavors over the past decades. To the best of my knowledge, at preset there are two different perspectives to solve this mystery, one is to resort to modified gravity theory. This kind of viewpoint is based on the idea that the general relativity (GR) should  be modified at large scale to incorporate the observed effects of Universe acceleration. One representative of such theories is the  $f(R)$ gravity proposed in 1970 \cite{Buchdahl:1970ynr}, where $f(R)$ introduced to the Lagrangian of the theory stands for an arbitrary function in terms of Ricci scalar $R$, and the freedom of the choices of $f(R)$ function makes it possible to explain the accelerated expansion.  The another viewpoint  to understand the accelerating Universe is by introducing an effective field (dark energy) which can generate repulsive gravity. A field with such repulsion feature can be formulated as a fluid with negative pressure in GR, and the most famous example of this fluid is the cosmological constant $\Lambda$ which was first proposed by Einstein and has been an essential ingredient of the $\Lambda \mathrm{CDM}$ model. Besides the cosmological constant, the so called phantom scalar field can also serve as a possible description of the dark energy \cite{Caldwell:1999ew,Cai:2009zp}. By comparing with the observational data \cite{Hannestad:2005fg,WMAP:2008rhx}, it has claimed that the phantom field endowed with negative energy density distribution can indeed be used to  explain the acceleration of our Universe. The phantom field can appear in the Einstein-Maxwell-dilaton system where the sign of dilatonic kinetic term is flipped
to be negative \cite{Clement:2009ai}, and it also emerges in the in the study of ghost branes in the string theory \cite{Okuda2006}.
Although the phantom field  may lead to quantum instabilities \cite{Caldwell:1999ew,Cai:2009zp} which poses a challenge to the theory, the authors in \cite{Piazza2004,Nojiri:2003vn} claimed that these instabilities can be avoided and it consequently makes phantom field a candidate for the dark energy model from the theoretical side. Given the usefulness of phantom field,
it therefore motivates us to study the physics of phantom field, especially to probe the effects of phantom field  on the  background of black holes spacetime if black holes solutions can be found in this scenario, since the black holes in the Universe will be inevitably affected by dark energy.

Recently, a new spherically symmetrical black hole solution was obtained in Einstein-anti-Maxwell theory with cosmological constant, called anti-RN-(A)dS solution, or simply called phantom RN-(A)dS black holes solution \cite{Jardim:2012se}. The phantom nature of the charge possessed by this kind of black holes make them drastically differ from their counterparts, namely usual RN-(A)dS black holes. Hence it is of great interest to study the characteristics of such black holes. The investigation of thermodynamical aspects of this black hole system have been performed in several works. The thermodynamics of phantom RN-AdS was discussed in \cite{Jardim:2012se}, followed by  a further study of geometrothermodynamics of this phantom black holes in \cite{Quevedo:2016cge}. The authors in \cite{Mo:2018hav} studied the phase transition and heat engine efficiency of phantom AdS black holes. While in this paper, we would like to focus on the dynamical properties of phantom RN-dS black holes under the scalar field perturbation, which means we are interested in investigating its QNMs spectrum on the background in our consideration.

QNMs has versatile applications in the black holes physics. One of its basic utilization is that we can use QNMs to examine the stability of black holes spacetime under perturbation \cite{Berti:2009kk,Konoplya:2011qq}. In the context of astrophysics, QNMs is contained in the gravitational waves (GWs) in the ringdown phase of  the mergers of binary black holes system and plays an increasingly essential role in the contemporary gravitational waves  astronomy \cite{LIGOScientific:2016aoc,LIGOScientific:2017vwq}, due to the fact that it can be regarded as characteristic ``sound'' of black holes \cite{Nollert:1999ji} and serve as the basis of black holes spectroscopy. Therefore, in principle,  rich informations of the GWs sources and spacetime geometry can be revealed with the successful detection of GWs. This feature of QNMs marks its great importance in the research of gravitational physics. In addition, QNMs has also been used to test GR and the validity of the famous ``no-hair'' theorem of black holes \cite{Dreyer:2003bv,Berti:2005ys, Shi:2019hqa, Isi:2019aib}, constrain modified gravity theories \cite{Liu:2020ddo,Bao:2019kgt,Cano:2021myl,Blazquez-Salcedo:2016enn,Franciolini:2018uyq,Aragon:2020xtm,Liu:2020qia,Karakasis:2021tqx} and examine strong cosmic censorship conjecture \cite{Cardoso:2017soq,Liu:2019lon,Hod:2018dpx,Mo:2018nnu,Dias:2018ynt,Hod:2018lmi,Gwak:2018rba,Guo:2019tjy}, and some other interesting discussions of QNMs  in the asymptotically dS spacetime  can be found in \cite{Sarkar:2023rhp,Konoplya:2022xid,Konoplya:2022kld,Zhidenko:2003wq}.
Given the significance of QNMs introduced above, hence it is intriguing and meaningful to study QNMs when a new black hole solution is obtained.

The present work is organized as follows. In Section \ref{section 2}, we first give a brief introduction to phantom RN-dS black holes, and then analyze the horizon structure and work out the value range of the phantom charge. In Section \ref{section 3}, we will numerically calculate QNMs spectrum of massless neutral scalar field perturbation with two methods and analyze the numerical results. The last section is devoted to conclusions. Throughout this paper, we will work with units $G=c=1$.

\section{phantom RN-dS black holes}\label{section 2}
In this section, we would like to briefly review phantom RN-dS black holes first, and then pin down its parameter space in which at least two horizons are present. The action of this theory is given by \cite{Jardim:2012se}

\begin{equation}
	S=\int d^4x\sqrt{-g}(R+2\eta F_{\mu\nu}F^{\mu\nu}+2\Lambda),\label{action}
\end{equation}
where $R$ is the Ricci scalar, $\Lambda$ is cosmological constant, and $F_{\mu\nu}$ is the coupled vector field strength whose nature is characterized by constant $\eta$.	For $\eta=1$, $F_{\mu\nu}$ is just the electromagnetic field strength, while for $\eta=-1$ it stands for phantom vector field strength. The terminology ``phantom'' is used here since the energy density of the field is negative for $\eta=-1$. 

Based on action given in Eq.~\eqref{action}, we can get a spherically symmetrical black hole solution \cite{Jardim:2012se}

\begin{equation}
	ds^2=-f(r)dt^2+f(r)^{-1}dr^2+r^2(d\theta^2+\sin \theta^2 d\phi^2),
\end{equation}
where 
\begin{equation}
	f(r)=1-\frac{2M}{r}-\frac{\Lambda}{3}r^2+\eta \frac{q^2}{r^2}.
\end{equation}
Obviously, this is the well-known RN-(A)dS spacetime metric when $\eta=1$. In the present paper, we are interested in the less concerned case $\eta=-1$ with a positive cosmological constant $\Lambda$>0, i.e. phantom RN-dS black hole spacetime. For our convenience, we would like to define a charge parameter  $Q\equiv\eta q^2$. With the parameter replacement, the phantom RN-dS black hole metric function $f(r)$ takes the form

\begin{equation}
	f(r)=1-\frac{2M}{r}-\frac{\Lambda}{3}r^2+ \frac{Q}{r^2}, \quad Q<0. \label{eq3}
\end{equation}

\begin{figure}
\centering
\includegraphics[height=3in,width=4.5in]{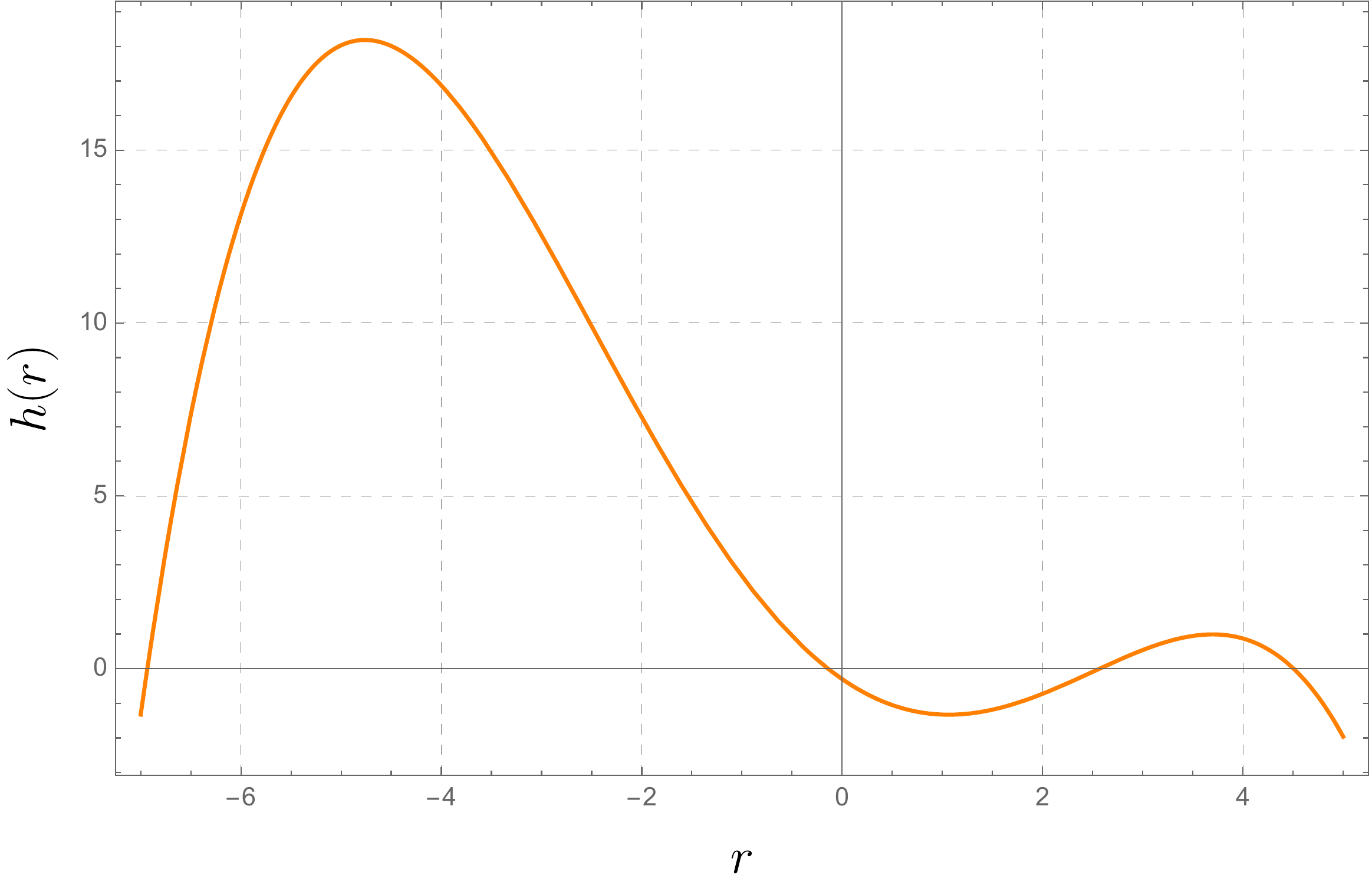}
\caption{The behavior of $h(r)$, where we take $M=1$, $\Lambda=0.08$ and $Q=-0.3$. Two negative and two positive roots are clearly present.}\label{fig1}
\end{figure}

To have a better understanding of this black hole spacetime, it is necessary to figure out the number of horizons, i.e. positive roots of function $f(r)$ of this spacetime. To simplify this task, we directly deal with function $h(r)\equiv r^2f(r)$ instead of dealing with $f(r)$. Apparently, the roots of $h(r)$ must be the roots of $f(r)$ except $r=0$, which is the location of black hole singularity so it is excluded from the possible roots. Function $h(r)$ is polynomial with the form
\begin{equation}
	h(r)=-\frac{\Lambda}{3}r^4+r^2-2Mr+Q,\label{funch}
\end{equation}
which is a quartic polynomial suggesting that there exists four roots for equation $h(r_i)=0$.  According to Vieta's theorem, the roots of $h(r)$ satisfy following relations
\begin{subequations}\label{vieta}
	\begin{align}
		&r_1+r_2+r_3+r_4=0,\\
		&r_1 r_2 r_3 r_4=-\frac{3Q}{\Lambda},
	\end{align}
\end{subequations}
where we have fixed black hole mass $M=1$ without loss of generality. For the physical reason, we only consider black hole parameters that yield four real roots $r_i\in\mathbb{R}$. Note that we have $\Lambda>0$ and $Q<0$ such that $r_1 r_2 r_3 r_4 =-3Q/\Lambda>0$. This means that the sign of $r_i$ has three different combinations, namely 1)four positive roots 2)two positive and two negative roots 3)four negative roots. On the other hand, we have $r_1+r_2+r_3+r_4=0$ which means that the roots can not be all positive  or all negative. Finally, with Eq.~\eqref{vieta}, we are forced to get two positive roots in addition to two negative roots, so there are at most two horizons in phantom RN-dS black hole spacetime. The two positive roots are labeled as $r_+$ and $r_c$ with the relation $r_+<r_c$. The function $h(r)$ in Eq.~\eqref{funch} can be rewritten as 
\begin{equation}
	h(r)=-\frac{\Lambda}{3}(r-r_1)(r-r_2)(r-r_+)(r-r_c),\label{funch2}
\end{equation}
where $r_1<r_2<0<r_+<r_c$. By Eq.~\eqref{funch2}, we can see that $h(r)<0$ in the region $0<r<r_+$ and $r>r_c$, whereas $h(r)>0$ in the region $r_+<r<r_c$. This fact reveals that $r_+$ and $r_c$ is event horizon and cosmological horizon, respectively. The behavior of $h(r)$ and the location of roots are clearly demonstrated in Fig.~\ref{fig1}, which shows the correctness of previous analysis.

Before taking a further investigation of the properties of phantom RN-dS black hole, we should figure out the parameter space in which both the event horizon and cosmological horizon can exist at the same time. Our first step is to find out the maximum  value of cosmological constant $\Lambda_{max}$, at which the minimum value of charge parameter $Q$ is zero. This extreme condition requires    

\begin{equation}
	Q=0, \quad h(r)=0,\quad h'(r)=0.
\end{equation}
The solution of equations above is
\begin{equation}
	\Lambda=\frac{3(r-2)}{r^3}=\frac{3(r-1)}{2r^3}.
\end{equation}
This equation  leads to $r=3$, which yields
\begin{equation}
	\Lambda_{max}=\frac{1}{9}.
\end{equation} 

\begin{figure}
\centering
\includegraphics[height=3in,width=4.5in]{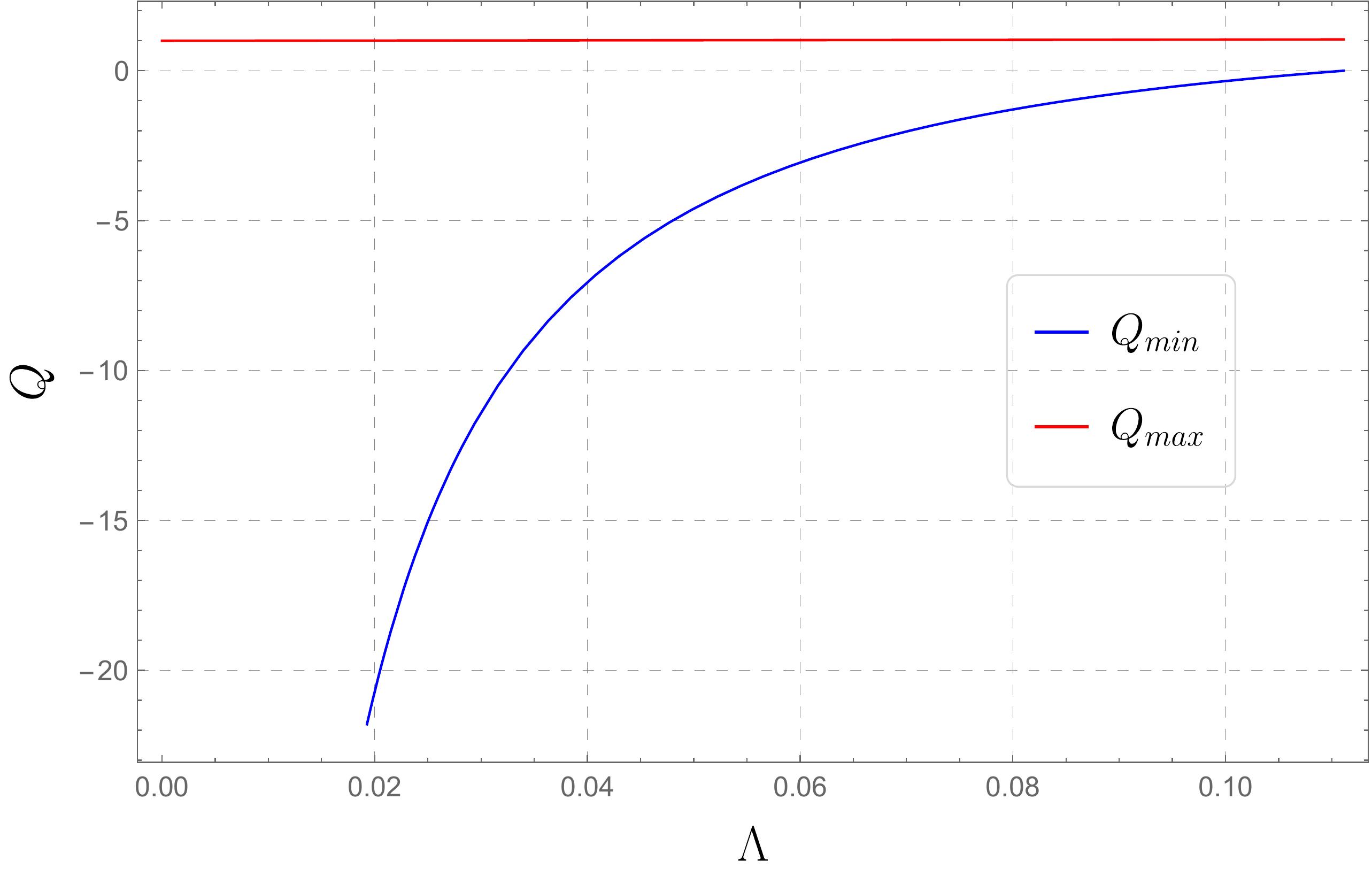}
\caption{The comparison between values of $Q_{min}$ and $Q_{max}$.}\label{fig2}
\end{figure}

For a given value of $\Lambda$, we are now supposed to calculate out the minimum and maximum value of $Q$, which is respectively denoted by $Q_{min}$ and $Q_{max}$. The similar strategy will be adopted as that used in finding $\Lambda_{max}$. After some simple calculation, we get analytical formula for $Q_{min}$ and $Q_{max}$ which are given by
\begin{subequations}
	\begin{align}
		&Q_{min}=-\frac{1}{\Lambda}\left(\cos\alpha+\sqrt{3}\sin\alpha\right)\left( -\sqrt{2\Lambda}+\frac{1}{4}\cos \alpha+\frac{1}{6}\cos3\alpha+\frac{\sqrt{3}}{4}\sin\alpha\right),\label{eq1} \\
		&Q_{max}=\frac{-2+\cos 2\alpha+3\sqrt{2\Lambda}\left(\cos \alpha-\sqrt{3}\sin \alpha\right)+ \sqrt{3}\sin 2\alpha
}{4\Lambda},	\end{align}
\end{subequations}
where we have defined 
\begin{equation}
	\alpha=\frac{1}{3}\arg \left(3\Lambda^2+\Lambda\sqrt{(9\Lambda-2)\Lambda}\right),
\end{equation}
in which  $\mathrm{arg}$  is the  argument function. We compare the values of $Q_{min}$ and $Q_{max}$ in terms of $\Lambda$ in Fig. \ref{fig2}, which shows that $Q_{min}$ is negative definite and and $Q_{max}$ is positive definite, and $Q_{max}$ is not as sensitive as $Q_{min}$ to the change of cosmological constant. This means that only $Q_{min}$ is related to charge parameter of  phantom RN-dS black hole, $Q_{max}$ is RN-dS black hole relevant. Actually, $Q_{min}$ marks the charge parameter value at which event horizon coincides with cosmological horizon, and $Q_{max}$ is only relevant to RN-dS black hole, since it is a positive  value where Cauchy horizon coincides with event horizon, while for phantom RN-dS black hole, it is simply required that $Q<0$ and no Cauchy horizon exists in this spacetime.
 When we set $\Lambda=\Lambda_{max}$, with Eq.~\eqref{eq1} we can directly get $Q_{min}=0$, this result is in agreement with our previous discussion about finding the value of $\Lambda_{max}$. It is well known that in RN-dS spacetime, charge parameter is limited to guarantee the existence of event horizon. In fact, the maxima and minima  of $Q_{max}$ respectively is
\begin{equation}
 \mathrm{Max}(Q_{max})=\lim\limits_{\Lambda\to\frac{2}{9}}Q_{max}=\frac{9}{8},\quad
 \mathrm{Min}(Q_{max})=\lim\limits_{\Lambda\to0^{+}}Q_{max}=1.
\end{equation}
Interestingly, on the contrary to RN-dS black hole, in phantom RN-dS black hole spacetime we find that $Q_{min}$ is not bounded from below, 
\begin{equation}
	\mathrm{Min}(Q_{min})=\lim\limits_{\Lambda\to0^{+}}Q_{min}=-\infty,
\end{equation}
which serves as  a remarkable difference between phantom and usual RN-dS black hole spacetime. 

\section{Quasinormal modes of the phantom RN-dS black holes}\label{section 3}
In this section, we focus on the calculation of QNMs frequencies of massless neutral scalar field perturbation  on the phantom RN-dS spacetime, and to this end we are required to derive the master equation of  scalar field perturbation. 

\subsection{Master Equation of Scalar Perturbation}
We start from a general static and spherically symmetrical  black hole metric in $3+1$ spacetimes,
\begin{equation}
ds^2=-A(r)dt^2+B(r)dr^2+r^2(d\theta^2+\sin^2\theta d\phi^2) .
\end{equation}
The equation of motion of a massless scalar field $\Phi$ reads
\begin{eqnarray}
\frac{1}{\sqrt{-g}}\partial_{\mu}(\sqrt{-g}g^{\mu\nu}\partial_{\nu}\Phi(t,r,\theta,\phi))=0.\label{eq5}
\end{eqnarray}
We decompose scalar field $\Phi(t,r,\theta,\phi)$ in terms of spherical harmonics function $Y_{lm}(\theta,\phi)$,
\begin{equation}
\Phi(t,r,\theta,\phi)=\sum_{l,m}\frac{\Psi(t,r)}{r}Y_{lm}(\theta,\phi)\label{eq4},
\end{equation}
where $l$ and $m$ stand for the angular and azimuthal number, respectively. Substituting Eq. \eqref{eq4} into Eq. \eqref{eq5}, we get the following radial equation
\begin{equation}
-\partial_t^2\Psi(t,r)+\frac{A}{B}\partial_r^2\Psi(t,r)+\frac{BA'-AB'}{2B^2}\partial_r\Psi(t,r)
+\frac{A(rB'-2l(l+1)B^2)-rBA'}{2r^2B^2}\Psi(t,r)=0,\label{eq6}
\end{equation}
where a prime denotes a derivative with respect to areal radius $r$. We simplify this equation by moving to tortoise coordinate $r_\ast$ defined by
\begin{equation}
dr_\ast=\sqrt{\frac{B(r)}{A(r)}}dr,
\end{equation}
Eq.~\eqref{eq6} can be rewritten as
\begin{equation}
-\frac{\partial^2\Psi(t, r)}{\partial t^2}+\frac{\partial^2\Psi(t,r)}{\partial r_\ast^2}-V(r)\Psi(t, r)=0 , \label{masterEqT}
\end{equation}
where the effective potential $V(r)$ is given by
\begin{equation}
V(r)=A(r)\frac{l(l+1)}{r^2}+\frac{1}{2r}\frac{d}{dr}\frac{A(r)}{B(r)}.\label{eqsc}
\end{equation}
We consider the time dependence of $\Psi(t,r)$ as $\Psi(t,r)=e^{-i\omega t}\phi(r)$, which gives rise to Schrodinger-like master equation,
\begin{equation}
\frac{d^2\phi(r)}{dr_\ast^2}+\left(\omega^2-V(r)\right)\phi(r) = 0 , \label{eq2}
\end{equation}
The black hole QNMs is determined by solving the eigenvalue problem defined by Eq.~\eqref{eq2} with the following boundary conditions for asymptotically de-Sitter and flat spacetimes 
\begin{equation}
\phi \sim
\begin{cases}
   e^{-i\omega r_\ast}, &  r_\ast \to -\infty, \\
   e^{+i\omega r_\ast}, &  r_\ast \to +\infty,
\end{cases}
\label{master_bc0}
\end{equation}
which indicate ingoing wave at the horizon and outoging wave at infinity.
Here, the eigenvalue $\omega$ is known as the QNMs frequency, which is usually a complex number due to the dissipative nature of boundary condition Eq.~\eqref{master_bc0} that makes the differential operator of this system non-self-joint.

It is time to get back to our specific phantom RN-dS spacetime metric. In our current case,  we have
\begin{equation}
A(r)=\frac{1}{B(r)}\equiv f(r),
\end{equation}
where metric function $f(r)$ is given by Eq.~\eqref{eq3} .
Accordingly, the effective potential Eq.~\eqref{eqsc} simplifies to read
\begin{equation}\label{eq7}
V(r)=f(r)\left(\frac{l(l+1)}{r^2}+\frac{f'(r)}{r}\right).
\end{equation}
With this effective potential and the master equation Eq.~\eqref{eq2} associated with boundary condition Eq.~\eqref{master_bc0}, the QNMs frequencies of scalar field perturbation in phantom RN-dS spacetime can be numerically obtained.

\subsection{Numerical Methods}
In this subsection, we calculate QNMs with the Asymptotic Iteration Method (AIM), of which an excellent review of this method can be found in \cite{Cho:2011sf}. At the same time, we also use  WKB approximation method which is improved by Pade approximants in order to verify the QNMs frequencies obtained by AIM. 

To employ AIM, we are required to deal with master equation Eq.~\eqref{eq2} in usual radial coordinate $r$, namely,
\begin{equation}
	f(r)f'(r)\phi'(r)+f^2(r)\phi''(r)+(\omega^2-V(r))\phi(r)=0. \label{eq8}
\end{equation}
We introduce a new variable  $\xi$,
\begin{equation}
	\xi=\frac{1}{r},
\end{equation}
and then take into consideration the asymptotical behavior (or boundary condition in Eq.~\eqref{master_bc0}) of $\phi(r)$, we rewrite $\phi(r)$ in terms of $\xi$ as follows,
\begin{equation}
\phi(\xi)=(\xi_+-\xi)^{-\frac{i\omega}{2\kappa_{+}}}(\xi-\xi_c)^{-\frac{i\omega}{2\kappa_{c}}}\chi(\xi)\label{eq9},
\end{equation}
where $\xi_+ =r_+^{-1}$ and $\xi_c=r_c^{-1}$, $\kappa_+=\frac{f'(r_+)}{2}$ and $\kappa_c=-\frac{f'(r_c)}{2}$ respectively are surface gravity on event horizon and cosmological horizon. To employ AIM, Eq.~\eqref{eq8} needs to be transformed into the form as
\begin{equation}
\chi''(\xi)=\lambda_{0}(\xi) \chi'(\xi)+s_{0}(\xi)\chi(\xi),
\end{equation}
where $\lambda_{0}$ and $s_{0}$ are given by 
\begin{equation}
\quad\lambda_0=\frac{-2\left(\Lambda+3 \xi^3(-M+Q \xi)\right)}{\xi\left(-\Lambda+3 \xi^2\left(1-2 M \xi+Q \xi^2\right)\right)}-\frac{-i \xi \omega \kappa_{+}\left(\xi-\xi_{+}\right)+\left(-i \xi \omega+2 \kappa_{+}\left(\xi-\xi_{+}\right)\right) \left(\xi-\xi_c\right)\kappa_c}{\xi \kappa_{+}\kappa_c\left(\xi-\xi_{+}\right) \left(\xi-\xi_c\right)},
\end{equation}

\begin{equation}
\begin{aligned}
&s_0=\frac{2 \Lambda-3 \xi^2\left(l+l^2+2 \xi(M-Q \xi)\right)}{\xi^2\left(\Lambda-3 \xi^2\left(1-2 M \xi+Q \xi^2\right)\right)}-\frac{9 \omega^2}{\left(\Lambda-3 \xi^2\left(1-2 M \xi+Q \xi^2\right)\right)^2}+ \\
& \qquad \frac{\omega}{4 \xi \kappa_{+}^2\left(\xi-\xi_{+}\right)^2 \kappa_c^2\left(\xi-\xi_c\right)^2}\left\{\xi \omega \kappa_{+}^2\left(\xi-\xi_{+}\right)^2+\left(\xi \omega+2 i \kappa_{+}\left(\xi-2 \xi_{+}\right)\right) \kappa_c^2\left(\xi-\xi_c\right)^2+\right. \\
& \qquad\left.2 \kappa_{+}\left(\xi-\xi_{+}\right) \kappa_c\left(i \kappa_{+}\left(\xi-\xi_{+}\right)\left(\xi-2 \xi_c\right)+\xi \omega\left(\xi-\xi_c\right)\right)\right\}+ \\
& \qquad\frac{i\omega\left(\Lambda+3 \xi^3(-M+Q \xi)\right) \left(\kappa_{+}\left(\xi-\xi_{+}\right)+\kappa_c\left(\xi-\xi_c\right)\right)}{\xi\left(-\Lambda+3 \xi^2\left(1-2 M \xi+Q \xi^2\right)\right)\left(\xi-\xi_{+}\right) \left(\xi-\xi_c\right)\kappa_{+}\kappa_c}
\end{aligned}
\end{equation}

Besides the numerical methods, WKB approximation method as a semi-analytical formula is also a powerful approach for finding QNMs frequencies. For our spherically symmetric background, the WKB formula gives a closed form of QNMs frequencies \cite{Konoplya:2019hlu},
\begin{equation}
\begin{aligned}
\omega^2&=V_0+A_2\left(\mathcal{K}^2\right)+A_4\left(\mathcal{K}^2\right)+A_6\left(\mathcal{K}^2\right)+\ldots\\
&-i \mathcal{K} \sqrt{-2 V_2}\left(1+A_3\left(\mathcal{K}^2\right)+A_5\left(\mathcal{K}^2\right)+A_7\left(\mathcal{K}^2\right) \ldots\right),
\end{aligned}\label{eq10}
\end{equation}
where $V_0$ is the value of effective potential at its maximum $V_0=V(r_{\ast 0})$, and so $r_{\ast 0}$ represents the location of the peak of $V(r_\ast)$. $V_2$ stands for the value of second order derivative of $V(r_\ast)$ respect to tortoise coordinate $r_\ast$ at the potential peak $r_{\ast 0}$. Henceforth, we simply denote the $m$-th  order derivative of $V(r_\ast)$  at $r_{\ast 0}$ as $V_{m}$,
\begin{equation}
V_{m}=\left.\frac{d^mV(r_\ast)}{dr_\ast^m}\right|_{r_\ast=r_{\ast 0}},\quad m\geq2.
\end{equation}
Obviously, for $m=1$ we have $V_1=0$. $A_{k}(\mathcal{K}^2)$ are polynomials of $V_2,V_3,\ldots V_{2k}$, and each $A_{k}(\mathcal{K}^2)$ should be considered as the $k$-th order corrections to the eikonal formula
\begin{equation}
\mathcal{K}=i\frac{\omega^2-V_0}{\sqrt{-2V_2}},	
\end{equation}
which provides unique solution for $\mathcal{K}$ with a given $\omega$. With the boundary conditions of QNMs, $\mathcal{K}$ is constrained to be 
\begin{equation}
\mathcal{K}=n+\frac{1}{2},\quad n\in\mathbb{N},	
\end{equation}
in which $n$ is the overtone number. With the given formula of $\mathcal{K}$ and Eq.~\eqref{eq10}, we are able to calculate QNMs frequencies directly. Here we list second and third order corrections as follows,

\begin{equation}
	A_2(\mathcal{K}^2)=\frac{-60\left(n+\frac{1}{2}\right)^2 V_3^2+36\left(n+\frac{1}{2}\right)^2 V_2 V_4-7 V_3^2+9 V_2 V_4}{288 V_2^2},
\end{equation}

\begin{equation}
\begin{aligned}
A_3(\mathcal{K}^2)&=\frac{1}{13824 V_2^5}\Bigg[-940\left(n+\frac{1}{2}\right)^2 V_3^4+1800\left(n+\frac{1}{2}\right)^2 V_2 V_4 V_3^2-672\left(n+\frac{1}{2}\right)^2 V_2^2 V_5 V_3\\
& -204\left(n+\frac{1}{2}\right)^2 V_2^2 V_4^2+96\left(n+\frac{1}{2}\right)^2 V_2^3 V_6-385 V_3^4+918 V_2 V_4 V_3^2-456 V_2^2 V_5 V_3 \\
&  -201 V_2^2 V_4^2+120 V_2^3 V_6\Bigg].
\end{aligned}
\end{equation}
For higher order corrections one can refer to  \cite{Konoplya:2019hlu} and references therein.

In order to increase the accuracy of higher order WKB method, the Pade approximants have been proposed to use in usual WKB formula \cite{Matyjasek:2017psv}. This approach is started by defining a polynomial $P_k(\epsilon)$ \cite{Konoplya:2019hlu},
\begin{equation}
\begin{aligned}
P_k(\epsilon)&=V_0+A_2\left(\mathcal{K}^2\right)\epsilon^2+A_4\left(\mathcal{K}^2\right)\epsilon^4+A_6\left(\mathcal{K}^2\right)\epsilon^6+\ldots\\
&-i \mathcal{K} \sqrt{-2 V_2}\left(\epsilon+A_3\left(\mathcal{K}^2\right)\epsilon^3+A_5\left(\mathcal{K}^2\right)\epsilon^5+A_7\left(\mathcal{K}^2\right)\epsilon^7 \ldots\right),
\end{aligned}\label{eq10}
\end{equation}
where the polynomial order $k$ is the same as the order of WKB formula. When $\epsilon=1$, one can get 
\begin{equation}
\omega^2=P_k(1).	
\end{equation}
The Pade approximants $P_{\widetilde{n}/\widetilde{m}}(\epsilon)$ for $P_k(\epsilon)$ near $\epsilon=0$ can be constructed as
\begin{equation}
	P_{\widetilde{n}/\widetilde{m}}(\epsilon)=\frac{Q_0+Q_1\epsilon+\ldots+Q_{\widetilde{n}}\epsilon^{\widetilde{n}}}{R_0+R_1\epsilon+\ldots+R_{\widetilde{m}}\epsilon^{\widetilde{m}}},
\end{equation}
where $\widetilde{n}+\widetilde{m}=k$, and $P_{\widetilde{n}/\widetilde{m}}(\epsilon)-P_{k}(\epsilon)=\mathcal{O}(\epsilon^{k+1})$.

\subsection{QNMs Frequencies of Phantom Black Holes}
In this subsection, we demonstrate fundamental(overtone number $n=0$)  scalar field  QNMs spectrum obtained by AIM and WKB method in Table \ref{tab1} and Table \ref{tab2}, and discuss the properties of these frequencies.

In Table \ref{tab1}, we display QNMs frequencies for different angular number $l$ and charge parameter $Q$, but cosmological constant is fixed to be $\Lambda=0.5\Lambda_{max}$. For each combination of parameters, the QNMs frequencies obtained by AIM and WKB method are putted together for comparison and cross-check. One can  notice that the results from our improved WKB method  and AIM are in great agreement with each other indicating the validity of these results, except for $l=0$, which reflects the well-known fact that WKB methods can give rise to reliable results only for QNMs with higher angular number $l>>n$. Although WKB formula corrected  by Pade approximation can greatly improve the accuracies for $l\sim n$ QNMs, we should still treat the QNMs frequencies with $l=n=0$ from WKB method carefully, as the accuracy for theses modes is not good enough. 
When  fixing  charge parameter and cosmological constant but increasing angular number $l$, one  can find that the real part of QNMs frequencies monotonously grow with $l$, as expected from the perspective of physics since the larger angular number corresponds to larger angular momentum which gives rise to a more rapid oscillation frequency. Whereas  for the imaginary part related to the damping rate of the modes, its magnitude decreases implying modes with larger $l$ will live longer.  On the other hand, when fixing angular number and cosmological constant but decreasing the charge parameter, we find that the real part of QNMs frequencies decreases while the imaginary part increases.

\begin{table}[!htbp]
\centering
\resizebox{\textwidth}{!}
{
        \begin{tabular}{cccccc}
    \hline\hline
    $l$ &Method& $Q=-0.5$         & $Q=-1$              & $Q=-2$                & $Q=-3$             \\
    \hline
    $0$ &AIM& $0.05545-0.09012i $ & $0.04506-0.08281i $ & $0.02739-0.06609i$    & $0.01134-0.04300i$\\
        \cline{2-6}
        &WKB& $0.04963-0.08836i $ & $0.03910-0.08021i $ & $0.02176-0.06209i$    & $0.00628-0.03887i$\\
        \hline
    $1$ &AIM& $0.16296-0.06517i $ & $0.13715-0.05728i $ & $0.09411-0.04140i$    & $0.05319-0.02389i$\\
        \cline{2-6}
        &WKB& $0.16299-0.06519i $ & $0.13717-0.05733i $ & $0.09406-0.04151i$    & $0.05312-0.02395i$\\
        \hline
    $2$ &AIM& $0.28331-0.06206i $ & $0.24115-0.05459i $ & $0.16906-0.03989i$    & $0.09726-0.02349i$\\
        \cline{2-6}
        &WKB& $0.28332-0.06206i $ & $0.24115-0.05459i $ & $0.16906-0.03989i$    & $0.09726-0.02349i$\\
        \hline
    $3$ &AIM& $0.40113-0.06130i $ & $0.34227-0.05396i $ & $0.24106-0.03954i$    & $0.13925-0.02340i$\\
        \cline{2-6}
        &WKB& $0.40113-0.06130i $ & $0.34227-0.05396i $ & $0.24106-0.03954i$    & $0.13925-0.02340i$\\
        \hline
    $5$ &AIM& $0.63468-0.06085i $ & $0.54233-0.05358i $ & $0.38296-0.03934i$    & $0.22175-0.02334i$\\
        \cline{2-6}
        &WKB& $0.63468-0.06085i $ & $0.54233-0.05358i $ & $0.38296-0.03934i$    & $0.22175-0.02334i$\\
        \hline
    $10$&AIM& $1.21573-0.06064i $ & $1.03955-0.05340i $ & $0.73497-0.03924i$    & $0.42606-0.02332i$\\
        \cline{2-6}
        &WKB& $1.21573-0.06064i $ & $1.03955-0.05340i $ & $0.73497-0.03924i$    & $0.42606-0.02332i$\\
    \hline\hline
\end{tabular}
}
\caption{The dominant QNMs frequency $\omega$ for $M=1$, $\Lambda=0.5\Lambda_{max}$, and $Q_{min} \approx -3.66843$.}\label{tab1}
\end{table}

In Table \ref{tab2}, we list fundamental QNMs frequencies for different cosmological constant $\Lambda$ while the  charge parameter is fixed to be $Q=-0.5$. We focus on the complex frequencies at the moment, and the same behavior can be observed as in Table \ref{tab1} when increasing angular number $l$, which leads to the real parts grow but the absolute value of imaginary parts decrease, and the WKB and AIM give rise to highly consistent results except for $l=0$. When we increase $\Lambda$, the real part of QNMs frequencies will decrease, while the imaginary parts increase. By observing the data in Table \ref{tab1} and Table \ref{tab2}, one can conclude that the increment of the magnitude of $Q$ or $\Lambda$ will results in the decrement of the magnitude of real and imaginary parts.

\begin{table}[!htbp]
\centering
\resizebox{\textwidth}{!}
{
        \begin{tabular}{ccccccccccc}
    \hline\hline
    $l$ &Method& $\Lambda=0.1\Lambda_{max}$& $\Lambda=0.3\Lambda_{max}$& $\Lambda=0.5\Lambda_{max}$ & $\Lambda=0.8\Lambda_{max}$ \\    \hline
    $0$ &AIM& $0.09501-0.10170i$ & $0.07769-0.09886i$ & $0.05545-0.09012i$    & $0.01242-0.04668i$\\
    \cline{2-6}
        &WKB& $0.09633-0.10256i$ & $0.07525-0.10043i$ & $0.04963-0.08836i$    & $0.01002-0.04134i$\\
        \hline
    $1$ &AIM& \makecell{$\boxed{0-0.06080i}$\\$0.25257-0.09100i$} & $0.21063-0.08053i$ & $0.16296-0.06517i$    & $0.06403-0.02572i$\\
    \cline{2-6}
        &WKB& $0.25257-0.09100i$ & $0.21060-0.08052i$ & $0.16299-0.06519i$    & $0.06399-0.02573i$\\
        \hline
    $2$ &AIM& $0.42016-0.08902i$ & $0.35700-0.07722i$ & $0.28331-0.06206i$    & $0.11588-0.02537i$\\
    \cline{2-6}
        &WKB& $0.42016-0.08902i$ & $0.35700-0.07722i$ & $0.28332-0.06206i$    & $0.11588-0.02537i$\\
        \hline
    $3$ &AIM& $0.58820-0.08844i$ & $0.50246-0.07633i$ & $0.40113-0.06130i$    & $0.16554-0.02529i$\\
    \cline{2-6}
        &WKB& $0.58820-0.08844i$ & $0.50246-0.07633i$ & $0.40113-0.06130i$    & $0.16554-0.02529i$\\
        \hline
    $5$ &AIM& $0.92440-0.08806i$ & $0.79226-0.07579i$ & $0.63468-0.06085i$    & $0.26329-0.02524i$\\
    \cline{2-6}
        &WKB& $0.92440-0.08806i$ & $0.79226-0.07579i$ & $0.63468-0.06085i$    & $0.26329-0.02524i$\\
        \hline
    $10$&AIM& $1.76490-0.08788i$ & $1.51507-0.07552i$ & $1.21573-0.06064i$    & $0.50559-0.02521i$\\
    \cline{2-6}
        &WKB& $1.76490-0.08788i$ & $1.51507-0.07552i$ & $1.21573-0.06064i$    & $0.50559-0.02521i$\\
    \hline\hline
\end{tabular}
}
\caption{The dominant QNMs frequency $\omega$ for $M=1$, $Q=-0.5$.}\label{tab2}
\end{table}

In Table \ref{tab1} and Table \ref{tab2}, the complex QNMs frequencies are classified into photon sphere modes (PS modes) \cite{Cardoso:2017soq}. Photon sphere is defined as the circular unstable geodesic trajectories that null particles are trapped on. The nearby area of this region is deeply connected to this kind of QNMs, as the authors in \cite{Cardoso:2008bp}  have found that black hole QNMs in the eikonal limit in any dimensions are determined by the parameters of the circular null geodesics on photon sphere, while there are claims that this correspondence is not perfectly guaranteed \cite{Konoplya:2022gjp,Konoplya:2017wot}. The dominant PS modes correspond to large $l$ limit and $n=0$, and they are well described by WKB approximation, as what we have shown in above two tables.

Among the complex PS modes, in Table \ref{tab2} one can notice a purely imaginary frequency in a black box for $\Lambda=0.1\Lambda_{max}$, $l=1$. This kind of modes come from the  memory of the pure dS spacetime (i.e. empty dS spacetime), so they are dubbed black hole dS modes (dS modes for short) which are deformations of the pure dS modes and are identified first in \cite{Jansen:2017oag} for neutral black hole spacetime and then confirmed for RN-dS spacetime in  \cite{Cardoso:2017soq}. In pure dS spacetime, the pure dS modes can be analytically expressed as \cite{Lopez-Ortega:2012xvr}
\begin{equation}
	\frac{\omega_{0, pure\,dS}}{\kappa_c^{dS}}=-il, \quad \frac{\omega_{n\neq0, pure\,dS}}{\kappa_c^{dS}}=-i(l+n+1).\label{eq11}
\end{equation}
The dominant dS modes ($l=1,n=0$) is almost identical to pure dS modes, but the deformations will grow for  modes with higher overtone number. Substituting $\Lambda=0.1\Lambda_{max}$ and $l=1$ into Eq.~\eqref{eq11} in which $\kappa_c^{dS}=\sqrt{\Lambda/3}$, we get pure dS modes frequency $\omega\approx -0.0608581 i$, which is very close to our dS modes frequency $\omega\approx-0.06080i$, and this coincidence proves that this frequency indeed belongs to the dS modes.  This modes only appear for $\Lambda=0.1\Lambda_{max}$ as a consequence of the fact that dS modes are dominant for $\Lambda\lesssim\Lambda_{cri}$. In usual RN-dS spacetime, it is claimed that  $\Lambda_{cri}\approx0.02$ \cite{Cardoso:2017soq} and it  seems also to be applicable in our case; on the other hand, in \cite{Cardoso:2017soq} it has also demonstrated that the fundamental dS modes is surprisingly weak dependent on black hole charge and they are almost identical to corresponding pure dS modes. While in our phantom RN-dS black hole spacetime, we will show that the critical value of $\Lambda_{cri}$ is dependent on the black hole charge parameter $Q$, meanwhile  the dS modes frequencies will be noticeably deformed by  $Q$ when it is big enough, as a consequence of $-Q$ can be arbitrarily large, but it will still remain a weak dependence on $Q$.

\begin{figure}[thbp]
\centering
\includegraphics[height=2.4in,width=3.2in]{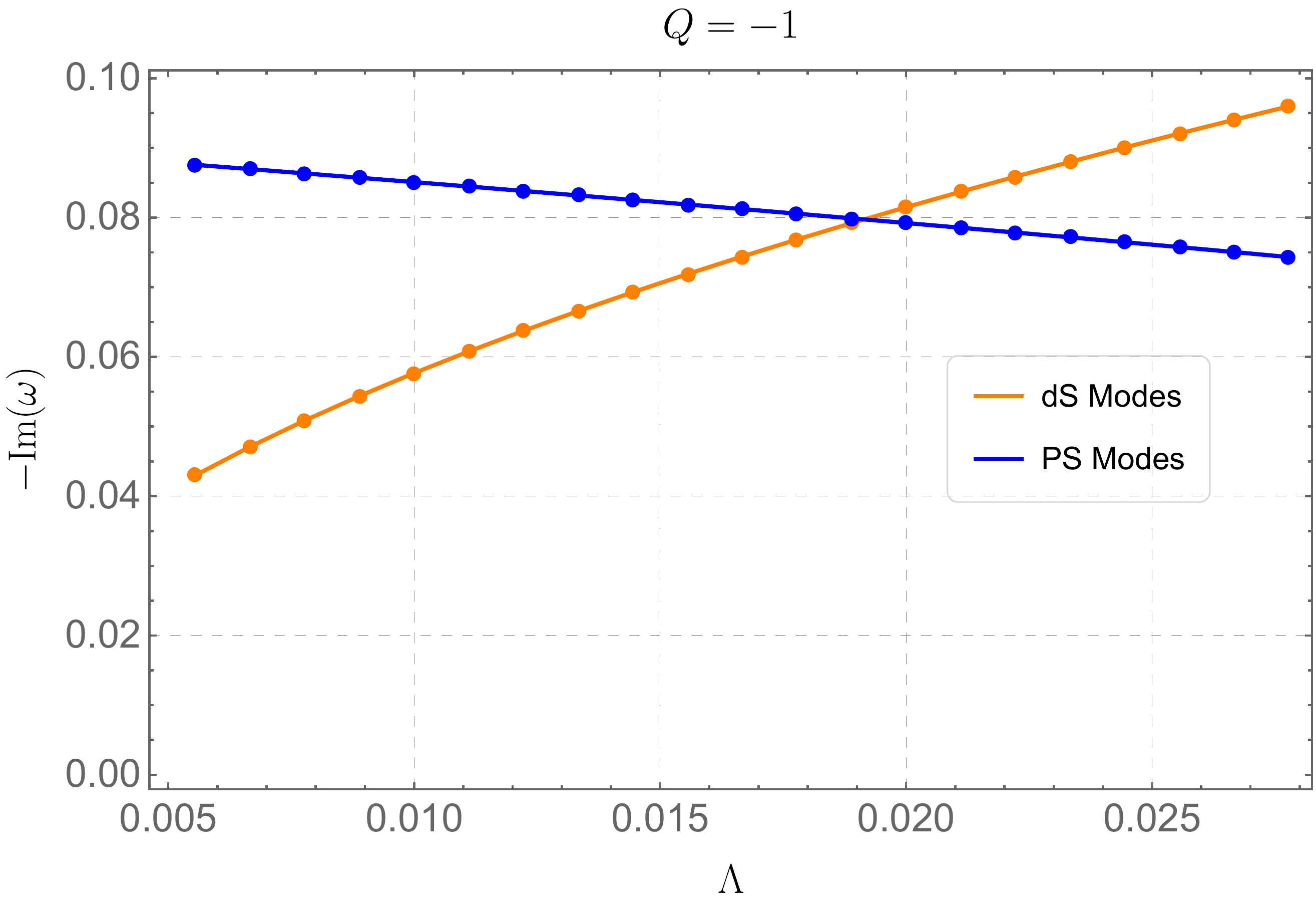}
\includegraphics[height=2.4in,width=3.2in]{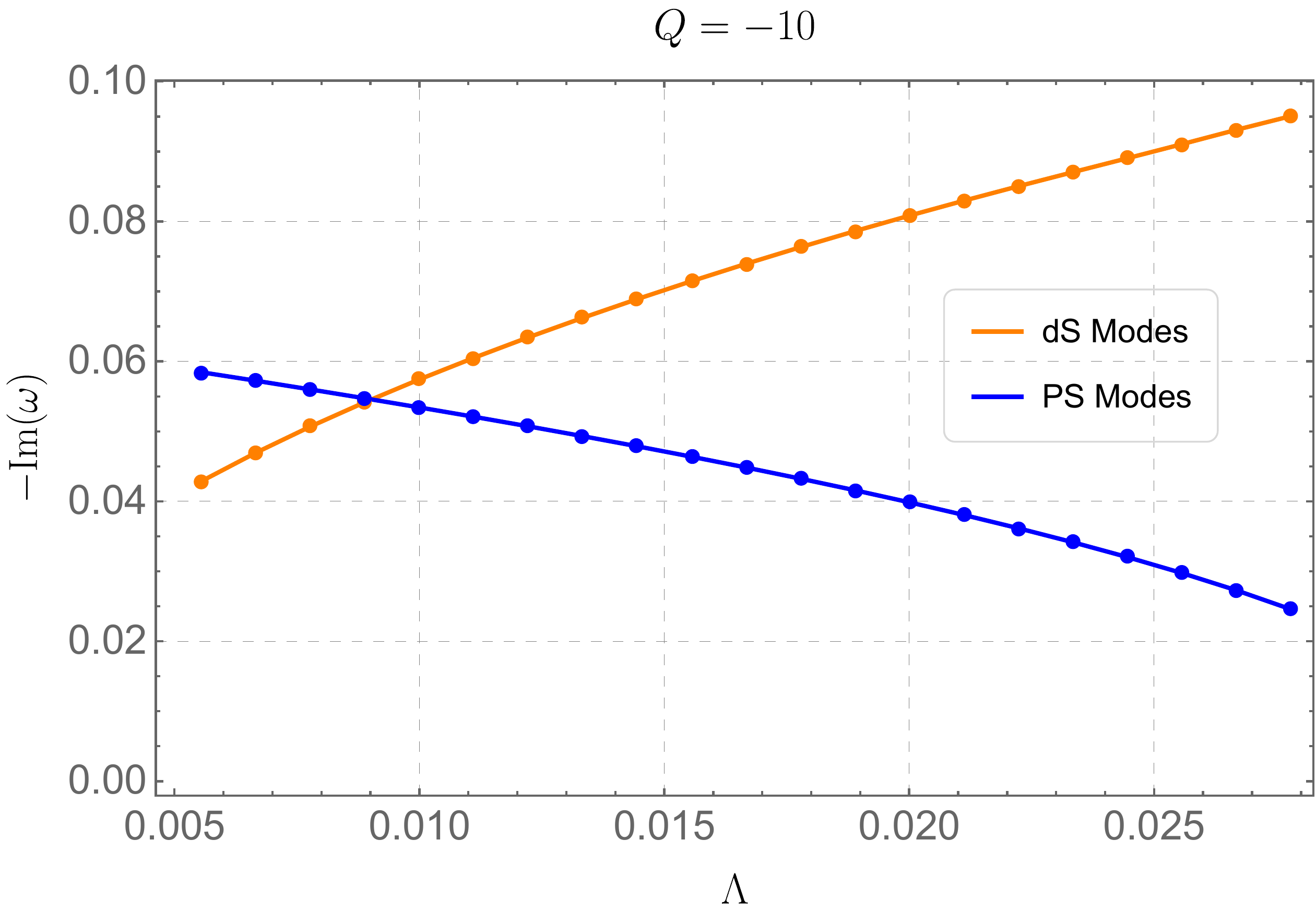}
\caption{The dependence of imaginary part of dominant PS modes (at large $l$ limit) and dominant dS modes ($l=1$) on the $\Lambda$ for $M=1$. The left panel is plotted for $Q=-1$, and the right panel corresponds to $Q=-10$.\label{fig3}}
\end{figure}

In Fig.~\ref{fig3}, we show the behavior of the $-\mathrm{Im}(\omega)$ of the  dominant PS and dS modes under the change cosmological constant $\Lambda$ for $Q=-1$ and $Q=-10$. One can observe that with the increase of $\Lambda$, the value of $-\mathrm{Im}(\omega)$ for dS  and PS modes behaves oppositely, that is  dS modes monotonously increase and PS modes monotonously decrease. A crosspoint can be identified in both plots and the corresponding value of $\Lambda$ is denoted as $\Lambda_{cri}$. When $\Lambda<\Lambda_{cri}$, the dS modes will dominate over PS modes as in this region dS modes have smaller $-\mathrm{Im}(\omega)$. 
For $Q=-1$, we find that $\Lambda_{cri}$ is about $\Lambda_{cri}\approx0.0189$, while when $Q$ is decreased to $-10$ one can find that $\Lambda_{cri}\approx0.0089$, which means that a smaller charge parameter $Q$ leads to a smaller $\Lambda_{cri}$.

\begin{figure}
\centering
\includegraphics[height=3in,width=4.5in]{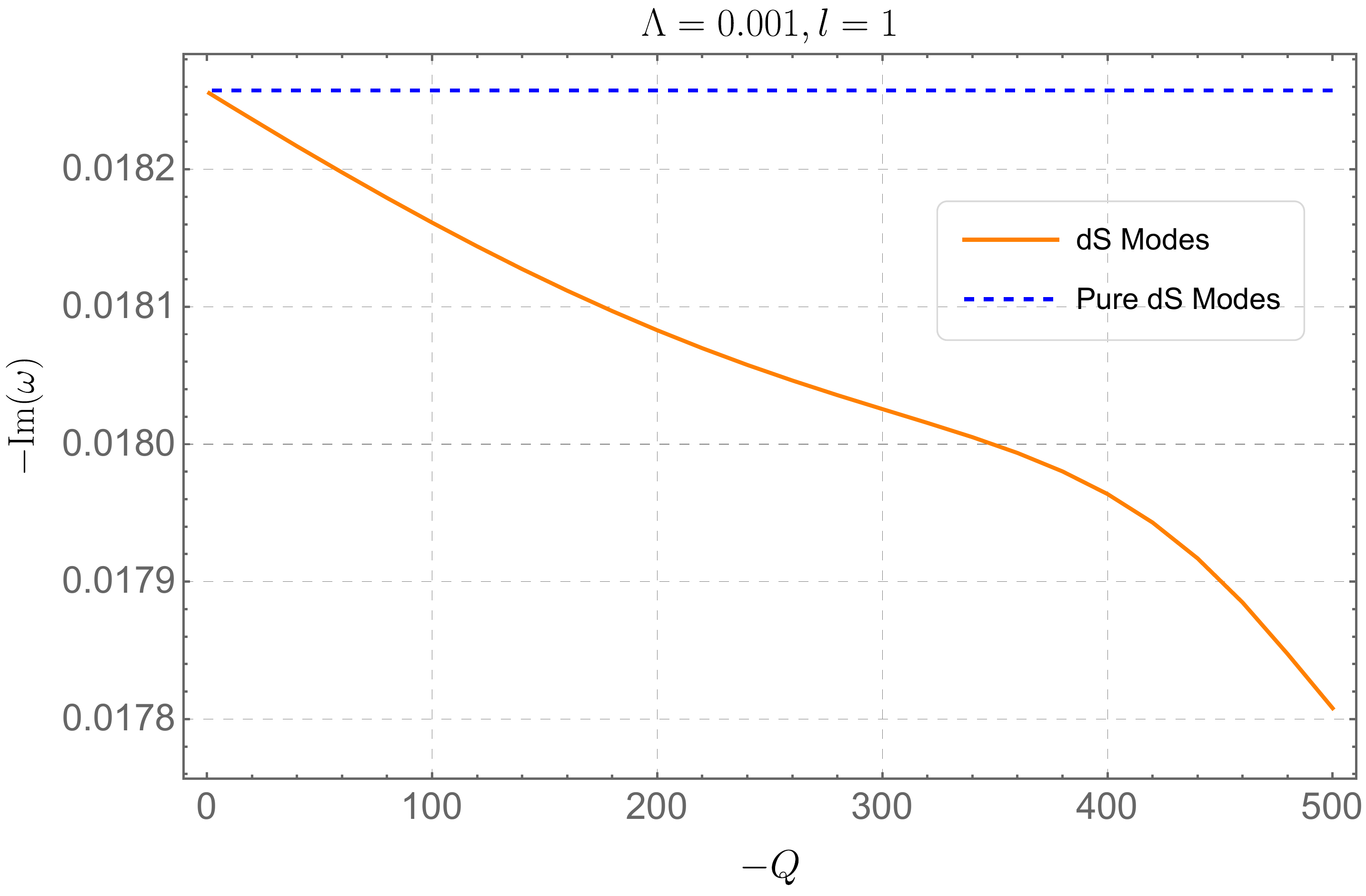}
\caption{The comparison between dominant dS modes and dominant pure dS modes with $M=1$.}\label{fig4}
\end{figure}

In Fig.~\ref{fig4}, we show the comparison of the value $-\mathrm{Im}(\omega)$ between dominant dS modes and dominant pure dS modes under the change of charge parameter $-Q$. The dominant pure dS modes depend only on cosmological constant $\Lambda$, given by $\omega=-\sqrt{\Lambda/3}i$, so it is a constant when $\Lambda$ is fixed, as shown in this figure. We find that dS modes is always more dominant than pure dS modes. When $-Q$ is close to zero, this two modes almost coincide with each other, but with the grow of $-Q$, the dS modes will gradually deviate from the pure dS modes, manifest as a larger $-Q$ results in a larger deviation. Although the deviation is noticeable and can be as large as $\sim0.0004$, it still remains a  quite weak  dependence  on charge. On the other hand, from the data listed in Table.~\ref{tab3}, we find that the frequency deviation from the pure dS modes is more sensitive to $\Lambda$, as the higher value of which can give rise to a greater frequency deviation, especially we have smaller $|Q|$ for bigger $\Lambda$.

\begin{table}[!htbp]
\centering
{
        \begin{tabular}{ccccccccccc}
    \hline\hline
     Modes Family   & \makecell{$\Lambda=0.01$\\$Q=-40$}& \makecell{$\Lambda=0.001$\\$Q=-500$}& \makecell{$\Lambda=0.0001$\\$Q=-6000$}\\    \hline
     $\omega_{\mathrm{pure\, dS}}$& $-0.057735i$ & $-0.0182574i$ & $-0.0057735i$\\
     \hline
     $\omega_{\mathrm{dS}}$& $-0.0566486i$ & $-0.0178078i$ & $-0.00560731i$\\
    \hline\hline
\end{tabular}
}
\caption{The dominant dS and pure dS modes ($l=1$) frequency $\omega$ at $M=1$.}\label{tab3}
\end{table}

\subsection{Comparison of QNMs Frequencies}
In this subsection, we would like to compare the QNMs frequencies between phantom and RN-dS black holes with the purpose of showing the differences between the two black holes and getting a further understanding of the effects of phantom charge on the QNMs. QNMs of different kinds of perturbation fields in the RN-dS spacetime have been extensively studied and one can find relevant calculations of QNMs in \cite{Cardoso:2017soq,Dias:2018etb,Mo:2018nnu,Dias:2018ufh}.

To make the comparison more natural, we will rewrite the metric function Eq.~\eqref{eq3} into following form,
\begin{gather}
	f(r)_{phantom}=1-\frac{2M}{r}-\frac{\Lambda}{3}r^2-\frac{Q}{r^2}, \quad Q>0,\\
	f(r)_{RN-dS}=1-\frac{2M}{r}-\frac{\Lambda}{3}r^2+\frac{Q}{r^2}, \quad Q>0,
\end{gather}
where $f(r)_{phantom}$ and $f(r)_{RN-dS}$ stands for the metric function for phantom and RN-dS black holes, respectively. In such form, we are able to compare the QNMs frequencies of the two black holes under the same value of charge  $Q$, as well as the other left parameters.

\begin{figure}[thbp]
\centering
\includegraphics[height=2.4in,width=3.2in]{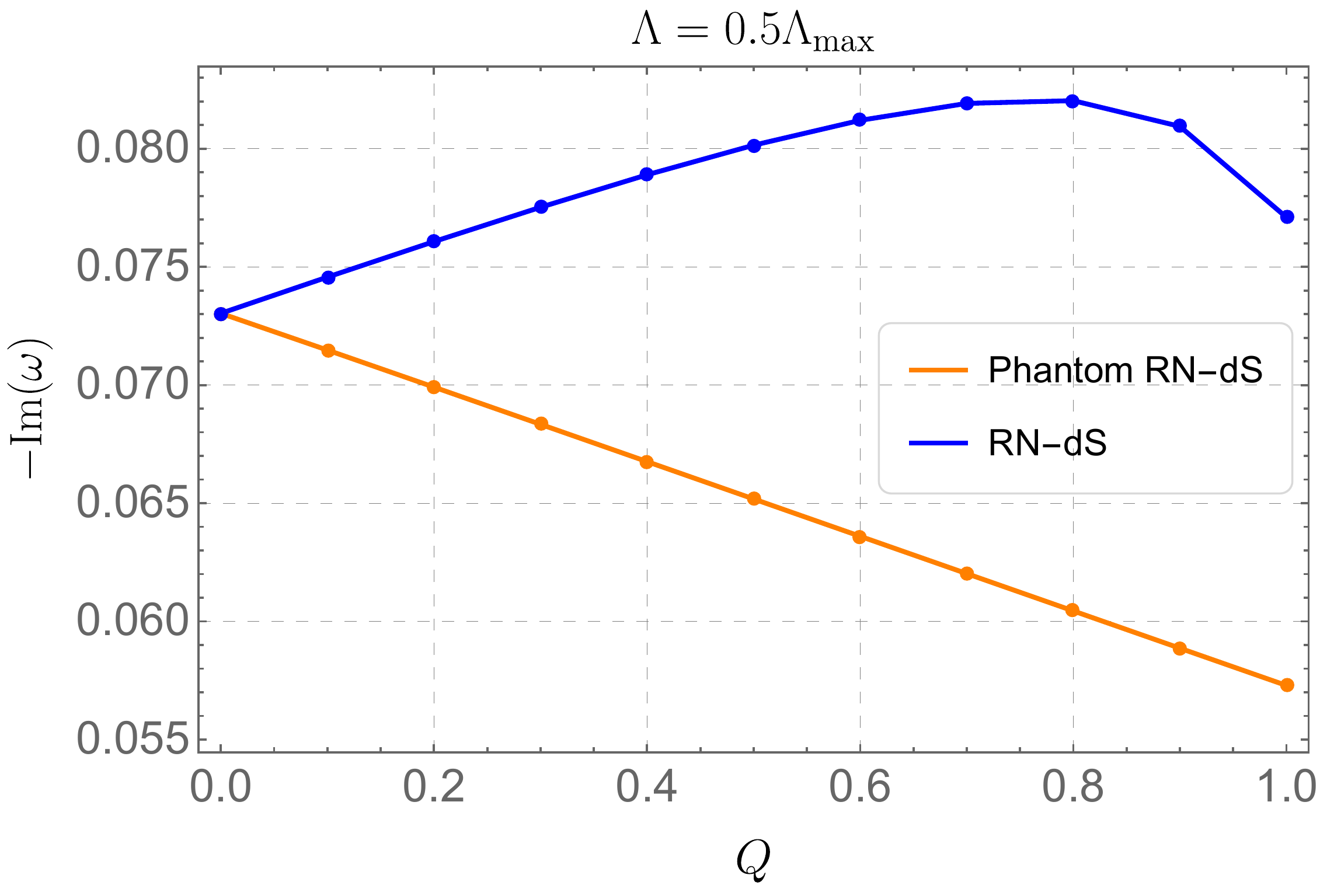}
\includegraphics[height=2.4in,width=3.2in]{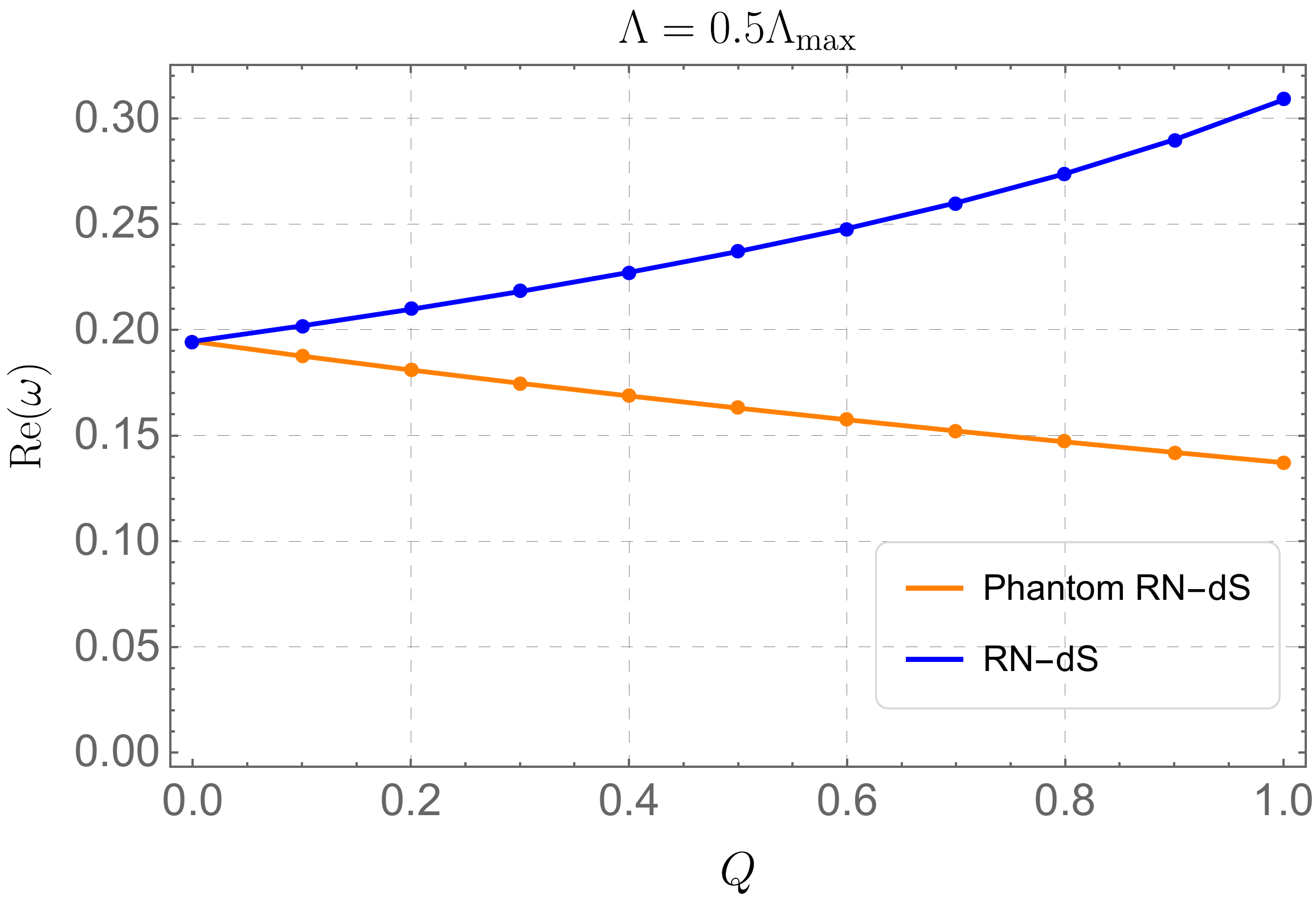}
\caption{The comparison of fundamental  QNNs frequencies of complex PS modes with $l=1$ between phantom and RN-dS black holes whose $M=1$ and $\Lambda=0.5\Lambda_{\mathrm{max}}$. The left plot and the right plot shows the imaginary part  $-\mathrm{Im}\,(\omega)$ and real part $\mathrm{Re}\,(\omega)$ of QNMs frequencies as a function of charge $Q$, respectively.\label{fig5}}
\end{figure}

In Fig.~\ref{fig5}, we separately  compare the frequencies of fundamental complex PS modes with angular number $l=1$ between phantom and RN-dS black holes. The comparison is made by varying charge value but other parameters remain unchanged. On the left panel, we can see that the imaginary frequency magnitude $-\mathrm{Im}\,(\omega)$ of RN-dS black holes is aways bigger than that of phantom RN-dS black holes, which means that QNMs will decay faster in RN-dS spacetime. When increasing charge, the imaginary frequency of phantom RN-dS black holes will monotonously and almost linearly decrease, indicating a larger charge will make the modes live longer. While for the RN-dS black holes, one can find that with the increase of charge the magnitude of imaginary frequency will grow until charge gets to around $\thicksim 0.8$ and then it will start to decrease. On the right panel, the real part of QNMs behaves in totally contrary way, i.e. a lager charge can make  $\mathrm{Re}\,(\omega)$ for RN-dS black holes bigger while  smaller for phantom RN-dS black holes, which leads to the differences between real frequencies of the two black holes monotonously increase with the charge. One can observe that $\mathrm{Re}\,(\omega)$ from RN-dS black holes is never smaller than phantom RN-dS black holes, which implies a more rapid oscillation frequency of QNMs in RN-dS spacetime.

\begin{figure}[thbp]
\centering
\includegraphics[height=2.4in,width=3.2in]{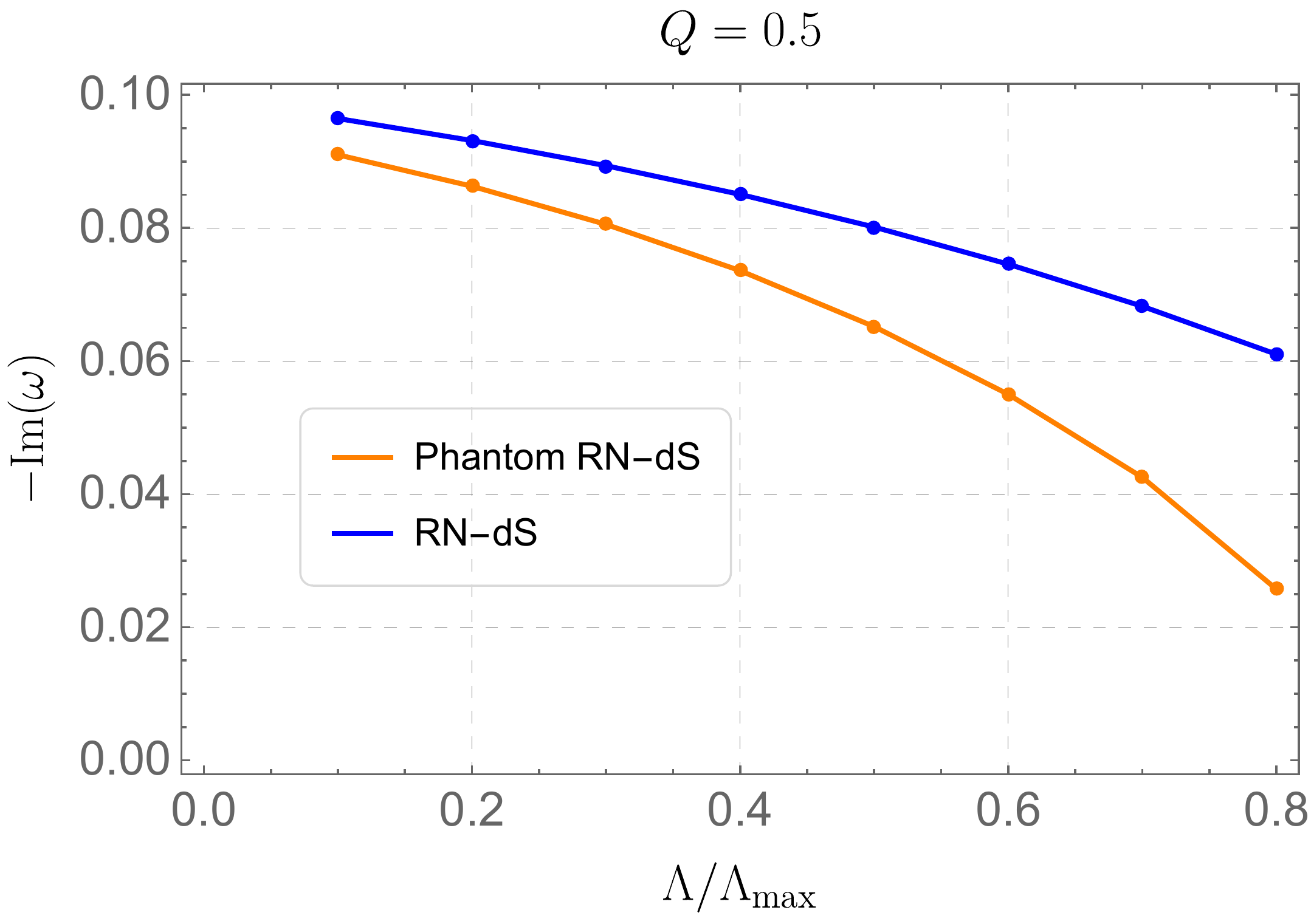}
\includegraphics[height=2.4in,width=3.2in]{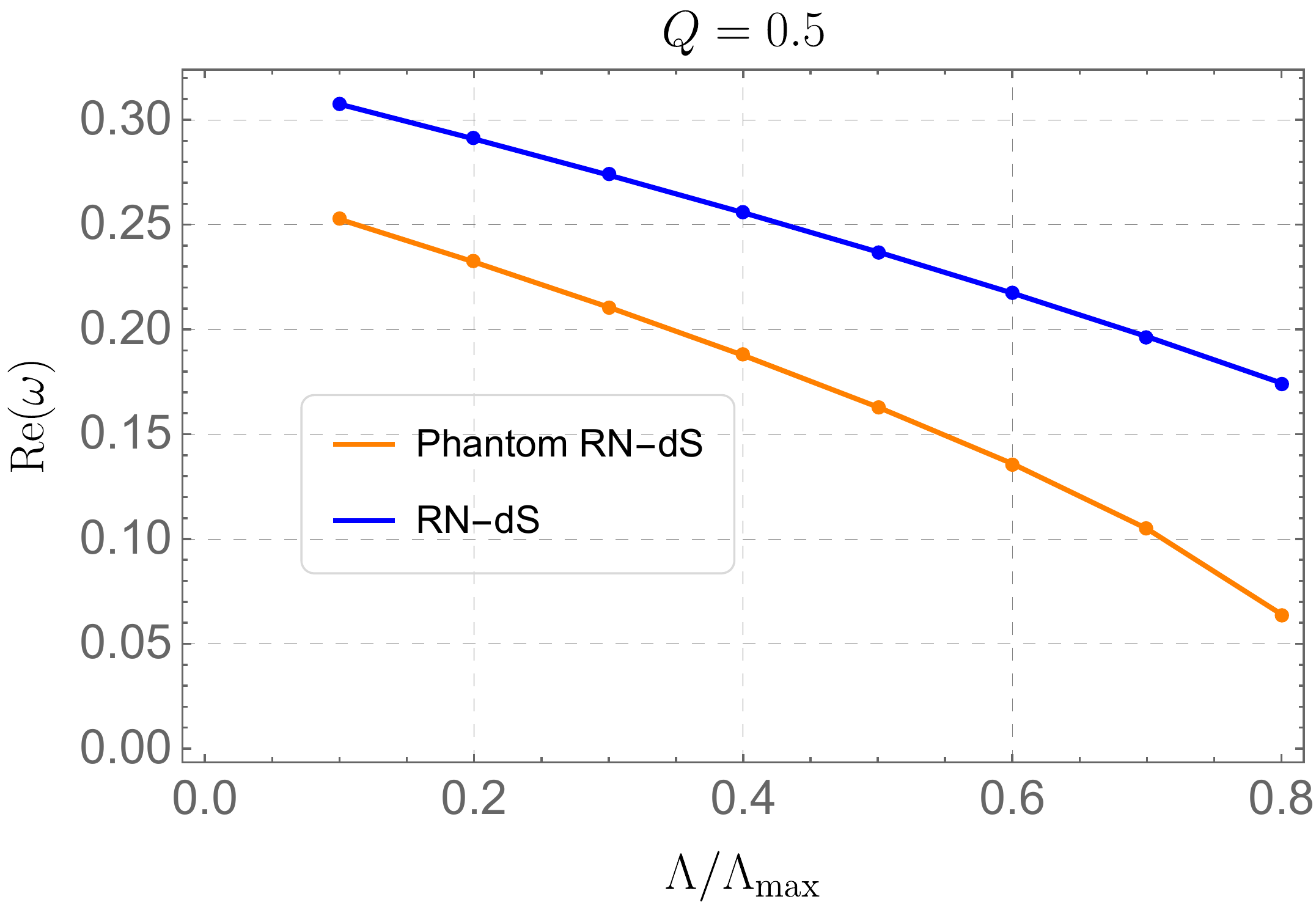}
\caption{The comparison of fundamental  QNNs frequencies of complex PS modes with $l=1$ between phantom and RN-dS black holes whose $M=1$ and $Q=0.5$. The left plot and the right plot shows the imaginary part  $-\mathrm{Im}\,(\omega)$ and real part $\mathrm{Re}\,(\omega)$ of QNMs frequencies as a function of $\Lambda/\Lambda_{\mathrm{max}}$, respectively.\label{fig6}}
\end{figure}

A similar comparison of QNMs frequencies of phantom and RN-dS black holes is demonstrated in Fig.~\ref{fig6} where the frequency curve is plotted as a function of the ratio $\Lambda/\Lambda_{\mathrm{max}}$ instead of charge $Q$. Under different $\Lambda$, we can see that both the imaginary and real part of QNMs frequencies from RN-dS black holes is always  higher than phantom RN-dS black holes, and this behavior has also been observed in Fig.~\ref{fig5}. On the contrary to the curves in Fig.~\ref{fig5}, for phantom and RN-dS black holes, both the imaginary and real part of frequencies synchronously decrease when increasing  $\Lambda$, which suggests a larger cosmological constant will give rise to a smaller oscillation frequency and a slower decay rate.

\section{Conclusions}

In this article, we have studied some properties of phantom RN-dS black hole, including its horizon structure, the value domain of the  charge parameter, and the QNMs spectrum of massless neutral scalar field perturbation. One of the  features of the phantom RN-dS black hole that differs from usual RN-dS black hole is that charge parameter of the phantom hole is negative. Under the negative charge condition, we find that there exists at most two horizons in this spacetime, namely event horizon and cosmological horizon. Especially, the value range of the phantom  black hole charge is dependent on cosmological constant $\Lambda$ and found to be not bounded from below when $\Lambda\to 0$, which exhibits a remarkable difference from normal charged black holes whose charge value is limited in order to avoid naked singularity.

We have analyzed QNMs spectrum of scalar field perturbation obtained by AIM and confirmed by WKB  approach which is greatly improved by Pade approximants, and classified the QNMs into dS modes and PS modes. When  the angular number $l$ is increased, one  can find that the real part of QNMs frequencies monotonously grow with $l$, as expected from the perspective of physics since a larger angular number corresponds to larger angular momentum which gives rise to a more rapid oscillation frequency. Whereas  for the imaginary part related to the damping rate of the modes, its magnitude decreases implying modes with larger $l$ will live longer. When we fix angular number and cosmological constant but decreasing the charge parameter, we find that the real part of QNMs frequencies decreases while the imaginary part increases. On the other hand,  when we solely increase $\Lambda$ and leave other parameters unchanged, the real part of QNMs frequencies will decrease, while the imaginary parts increase. With these results, we  conclude that the increment of the magnitude of $Q$ or $\Lambda$ will result in the decrement of the magnitude of real and imaginary parts. The dS modes have the chance to become the dominant modes over PS modes, when $\Lambda<\Lambda_{cri}$. We find that $\Lambda_{cri}$ is related to the value of charge parameter $Q$, as for a larger $|Q|$, $\Lambda_{cri}$ will be smaller. Finally, we examine the deviations of dS modes from the pure dS modes. It is known that dS modes originate from pure dS modes, such that in a asymptotically dS black hole spacetime, the dominant dS modes frequencies are almost identical to the corresponding pure dS modes frequencies, just with a tiny deformation. We find that this deformation depends on charge parameter $Q$ and cosmological constant $\Lambda$. A larger $\Lambda$ or $|Q|$ can lead to a larger deviation, which can be noticeable and seem to be more sensitive to the variation of $\Lambda$. 

At last, we compared the QNMs frequencies of phantom and RN-dS black holes in order to reveal more effects of phantom charge and highlight the differences between the two black holes. We find that under any combinations of parameters in our consideration, both imaginary and real part of QNMs frequencies from RN-dS black holes are never smaller than that of phantom RN-dS black holes, which means that the QNMs of phantom RN-dS black holes can live longer and oscillate less rapidly compared to RN-dS black holes. When charge is fixed and increasing cosmological constant, we find that $-\mathrm{Im}\,(\omega)$ and  $\mathrm{Re}\,(\omega)$ will decrease, for both black holes. On the other hand, it was found that when the $\Lambda$ is fixed, for phantom RN-dS  black holes the $-\mathrm{Im}\,(\omega)$ and $\mathrm{Re}\,(\omega)$ will become smaller with the grow of $Q$. However, for RN-dS black holes a larger $Q$ will make $\mathrm{Re}\,(\omega)$ monotonously increase but  $-\mathrm{Im}\,(\omega)$ behave  non-monotonically.

\begin{acknowledgments}
My most sincere thanks go to  Yanfei for her consistent support to my career. This work is supported by the Natural Science Foundation of China under Grant No.12305071.
\end{acknowledgments}

\bibliographystyle{JHEP}
\bibliography{References_Phantom_BH}

\end{document}